\documentclass[aps, pre, reprint, superscriptaddress, showpacs, amsmath, amssymb]{revtex4-1}
\usepackage{graphicx}

\newcommand{\Eq}[1]{Eq.~(\ref{#1})}
\newcommand{\Fig}[1]{Fig.~(\ref{#1})}
\newcommand{\Dif}[0]{\textrm{d}}

\newcommand{\Cosh}[0]{\textrm{Ch} }

\begin{document}


\title{Self-Similar Solutions to a Density-Dependent Reaction-Diffusion Model}


\author{Waipot Ngamsaad}
\email{waipot.ng@up.ac.th}
\affiliation{Division of Physics, School of Science, University of Phayao, Mueang Phayao, Phayao 56000, Thailand}

\author{Kannika Khompurngson}
\affiliation{Division of Mathematics, School of Science, University of Phayao, Mueang Phayao, Phayao 56000, Thailand}
\affiliation{Centre of Excellence in Mathematics, PERDO, CHE, Thailand}


\date{\today}

\begin{abstract}
In this paper, we investigated a density-dependent reaction-diffusion equation, $u_t = (u^{m})_{xx} + u - u^{m}$. This equation is known as the extension of the Fisher or Kolmogoroff-Petrovsky-Piscounoff equation which is widely used in the population dynamics, combustion theory and plasma physics. By employing the suitable transformation, this equation was mapped to the anomalous diffusion equation where the nonlinear reaction term was eliminated. Due to its simpler form, some exact self-similar solutions with the compact support have been obtained. The solutions, evolving from an initial state, converge to the usual traveling wave at a certain transition time. Hence, it is quite clear the connection between the self-similar solution and the traveling wave solution from these results. Moreover, the solutions were found in the manner that either propagates to the right or propagates to the left. Furthermore, the two solutions form a symmetric solution, expanding in both directions. The application on the spatiotemporal pattern formation in biological population has been mainly focused.
\end{abstract}


\maketitle

\section{Introduction}
Many transport phenomena arising in natural science have been successfully described by the nonlinear reaction-diffusion (NRD) model. The applications of the NRD model have had wide variety including population dynamics, transport in porous medium, combustion theory and plasma physics \cite{Aronson1986, Samarskii1995, Barenblatt1996, Murray2002, Petrovskii2006}. The NRD equation is formulated by
\begin{equation}\label{eq:gen_RD}
u_t = [D(u)u_x]_{x} + R(u) ,
\end{equation} 
where $u := u(x,t) \geqslant 0$, depending on the system,  is density or temperature at position $x \in (-\infty, \infty)$ and time $t \in [0, \infty)$. $D(u)$ and $R(u)$ are the diffusion coefficient and the reaction term respectively. The simplest form of NRD \Eq{eq:gen_RD} is the well known Fisher equation \cite{Fisher1937}, as well as the Kolmogoroff-Petrovsky-Piscounoff (KPP) equation,  which was originally derived for the propagation of a gene in a population. In this classic model, the reaction term is represented by the self-limiting Pearl-Verhulst logistic law $R(u) = \alpha u(1-u/\sigma)$ and the diffusion is represented by the Brownian process, reflected through the term $D(u) = \kappa$, where $\kappa > 0$ is the diffusion constant, $\alpha > 0$ is rate constant and $\sigma = \displaystyle\lim_{t \to \infty} u(x,t)$ is equilibrium density. Since the Brownian motion seems to be not realistic movement for the active individual such as biological population, the density-dependent diffusion has been included which represented by $D(u) = \kappa (u/\sigma)^p$ where $p > 0$ \cite{Gurney1975, Gurney1976, Gurtin1977, Newman1980, Newman1983}. It reflects the crowd avoiding movement of the individual. Later, it has been also found the more general form for the logistic law $R(u) = \alpha u[1-(u/\sigma)^p]$ as well \cite{Newman1983}. By substituting $D(u)$ and $R(u)$ in the later case into \Eq{eq:gen_RD} with the transformations $t^\ast = \alpha t$, $x^\ast = (m \alpha /\kappa)^\frac{1}{2}x$ and $u^\ast = u/\sigma$, we obtain the dimensionless equation 
\begin{equation}\label{eq:gen_fisher_dimless}
u_t = (u^{m})_{xx} + u - u^{m} ,
\end{equation} 
where $m=p+1>1$ and the asterisk is dropped for convenience. The applications of this equation have been found in various systems (see \cite{Aronson1986, Samarskii1995, Barenblatt1996, Murray2002, Petrovskii2006} for instance). Due to the diffusion coefficient is depended on the density; this equation is so called the density-dependent reaction-diffusion (DDRD) equation. Sometime, this model is called the generalized or extended Fisher equation \cite{Murray2002}. Since the existence of the traveling wave solution in the NRD equation has been found in both the original Fisher equation \cite{Fisher1937} and the DDRD equation \cite{Newman1980, Newman1983}, the analysis on these equations has been attracted much attention (see  \cite{Murray2002, Petrovskii2006} for review). 

Recently, the NRD model has been employed to describe the spatiotemporal pattern formation in bacterial colonies which is the elegant self-organization observed in the biological system \cite{Kawasaki1997, Golding1998, Ben-Jacob2000}. For each species, the bacterial colonies generate various patterns depending on surrounding environments, mainly, including the nutrient concentration and the softness of the medium. This cooperative behavior in adaptation to survival in the environment reflects bacterial communication capabilities and social intelligence \cite{Golding1998, Ben-Jacob2004}. The understanding of the underlying mechanism is not only important to the biotechnology but also the basic science of the living organisms. Kawasaki \textit{et al.} \cite{Kawasaki1997} pointed out that although the bacterial colony evolves in two dimensions, each tip grows in one dimension except for occasional branching. Therefore, the velocity of the tip elongation obtained form the one-dimensional system could be compared well with one obtained from the two-dimensional system. They also proposed the simplified version of the model for pattern formation in bacterial colony which was consistent with \Eq{eq:gen_fisher_dimless} \cite{Kawasaki1997, Ben-Jacob2000}.

In this paper, we investigate the generalized or extended Fisher equation (\Eq{eq:gen_fisher_dimless}) in one-dimensional space \cite{Newman1983, Murray2002}. Although the asymptotic behavior of its solution has well understood as the traveling wave solutions from the past studies, the exact or event explicit solution of \Eq{eq:gen_fisher_dimless} has been unknown. That hinders the understanding how the systems develop to the traveling wave from the initial state. In different aspect, the self-similar solution has been substituted directly to \Eq{eq:gen_fisher_dimless} \cite{Biro1997}. It turned out that, after evaluating, the solution was difficult to get the closed-form. The analysis on the reduced form of \Eq{eq:gen_fisher_dimless} has been also carried out. In 1977, Gurtin and MacCamy  \cite{Gurtin1977} have shown that the NRD, with Malthusian law as the reaction term, can be reduced to the simpler form by the simple transformation, which the reaction term was eliminated. Thus, equation transforms to the purely nonlinear diffusion equation. In 1983, Newman \cite{Newman1983} has transformed \Eq{eq:gen_fisher_dimless} to the reduced form until the integral invariant and Lyapunov functional have been admissible. The convergence of the solutions toward to the traveling wave has been obtained as the asymptotic behavior. In 2002, Rosenau \cite{Rosenau2002} has employed a transformation which mapped \Eq{eq:gen_fisher_dimless} to  the purely nonlinear diffusion where the the nonlinear reaction term $R(u)$ was eliminated. According to the inhomogeneous nature of the mapped equation, the solution has been obtained only at asymptotic form and has not been in the closed-form. In 2004, Harris \cite{Harris2004} has also applied the similar method of Rosenau \cite{Rosenau2002} to analyze \Eq{eq:gen_fisher_dimless}. Indeed, the explicit solution, which evolves from the initial state to the traveling wave, has been obtained for $m=2$. However, the solution for the case $m>2$ has been unknown. 

In our analysis, we also follow the similar method of \cite{Rosenau2002, Harris2004}. But, the \Eq{eq:gen_fisher_dimless} is differently mapped to the anomalous diffusion equation which the exact self-similar solution is obtained for any $m>2$ \cite{Bologna2000, Tsallis2002, Lenzi2003}. It is important to point out that the connection between the self-similar solution to the traveling wave solution, as described in \cite{Barenblatt1972}, has seen clearly from this particular solution. The outlines of this paper are as followings. The details of analysis for the solution of \Eq{eq:gen_fisher_dimless} are described in Sec. \ref{sec:solutions}. The discussions on the obtained results are given in Sec. \ref{sec:discussions}. Finally, we conclude our study in Sec. \ref{sec:conclusions}.

\section{\label{sec:solutions}Main Results}
We now perform the analysis on \Eq{eq:gen_fisher_dimless}. Noticing that, \Eq{eq:gen_fisher_dimless} can be factorized as $(\frac{\partial}{\partial t} - 1)u = (\frac{\partial}{\partial x} + 1)(\frac{\partial}{\partial x} - 1)u^m$ and then it can be written in the equivalent equation
\begin{equation}\label{eq:tranfisher_1}
e^{t}\frac{\partial}{\partial t}e^{-t} u = e^{-x}\frac{\partial}{\partial x}e^{x}(e^{x}\frac{\partial}{\partial x}e^{-x} u^m) .
\end{equation} 
It suggests that \Eq{eq:tranfisher_1} can be reduced to the simpler form by seeking the solution in the composition of the exponential factors
\begin{equation} \label{eq:sol_1}
u(x,t) = e^t e^{\frac{x}{m}} \Phi(x,t) ,
\end{equation}
where $\Phi(x,t) \geqslant 0$ is the transformed density. By substituting \Eq{eq:sol_1} into \Eq{eq:tranfisher_1}, we obtain
\begin{equation}\label{eq:tranfisher_2}
e^{-(m-1)t} \Phi_t = e^{-\frac{m+1}{m}x} [ e^{2x} (\Phi^m)_x ]_x .
\end{equation}
To eliminate the exponential terms in \Eq{eq:tranfisher_2},  we define $(m-1)e^{(m-1)t}\Dif t = \Dif \tau $ and then the temporal transform-function is given by  
\begin{equation}\label{eq:time_1}
\tau(t) = e^{(m-1)t} - 1 ,
\end{equation}
which satisfies the initial condition $\tau(0)=0$. Similarly, we define $\frac{m+1}{m} e^{\frac{m+1}{m}x}\Dif x = \Dif \phi$ and then we evaluate the spatial transform-function  
\begin{equation}\label{eq:space_1}
\phi(x) =  e^{\frac{m+1}{m}x} .
\end{equation}
With transformations \Eq{eq:time_1} and \Eq{eq:space_1}, the intrinsic space and time are eliminated. 
Consequently, \Eq{eq:tranfisher_2} is reduced to
\begin{equation}\label{eq:ab_diff_1}
\Phi_\tau = \tfrac{1}{m-1} \left( \tfrac{m+1}{m} \right) ^ 2 [ \phi^{\frac{3m+1}{m+1}}  (\Phi^m)_{\phi} ]_{\phi} .
\end{equation}
\Eq{eq:ab_diff_1} is similar to the \textit{anomalous diffusion equation} $\Phi_\tau = [ k \phi^{l}  (\Phi^m)_{\phi} ]_{\phi}$, where $k = \frac{1}{m-1} \left( \frac{m+1}{m} \right) ^ 2$ and $l = \frac{3m+1}{m+1} > 2$, which the exact solution is known \cite{Bologna2000, Tsallis2002, Lenzi2003}. To this end, it can be interpreted that the nonlinear reaction-diffusion \Eq{eq:gen_fisher_dimless} is mapped to the anomalous diffusion \Eq{eq:ab_diff_1} by employing the transformations \Eq{eq:sol_1}, \Eq{eq:time_1} and \Eq{eq:space_1}. 

Similar to \cite{Bologna2000, Tsallis2002, Lenzi2003}, the solution of \Eq{eq:ab_diff_1} is assumed to be the normalized scaled function of the type 
\begin{equation}\label{eq:ab_diff_sol}
\Phi(\phi,\tau) := \frac{1}{T(\tau)}F\left(\frac{\phi}{T(\tau)}\right) = \frac{F(\omega)}{T(\tau)} ,
\end{equation}
where $\omega(\phi,\tau) := \phi/T(\tau)$. By substituting \Eq{eq:ab_diff_sol} into \Eq{eq:ab_diff_1}, we obtain
\begin{equation}\label{eq:ab_diff_2}
-\frac{T_\tau}{T^2}  ( F + \omega F_\omega ) = \tfrac{1}{m-1} \left( \tfrac{m+1}{m} \right)^2 \frac{[\omega^{\frac{3m+1}{m+1}} (F^m)_\omega]_\omega}{T^{2+m-\frac{3m+1}{m+1}}} . 
\end{equation}
\Eq{eq:ab_diff_2} can be separated into two equations; The first and the second equations are given by
\begin{equation}\label{eq:ab_diff_time_1}
T^{m-\frac{3m+1}{m+1}} T_\tau = \tfrac{m+1}{m(m-1)} \lambda^{\frac{m(m-1)}{m+1}} ,
\end{equation}
and
\begin{equation}\label{eq:ab_diff_func_1}
-\lambda^{\frac{m(m-1)}{m+1}} (\omega F)_\omega = \tfrac{m+1}{m} [\omega^{\frac{3m+1}{m+1}} (F^m)_\omega]_\omega ,
\end{equation}
respectively, where $\lambda$ is an arbitrary positive constant. According to \Eq{eq:ab_diff_time_1}, we obtain  the solution 
\begin{equation}\label{eq:ab_diff_time_2}
T(\tau) = \lambda (\tau + a )^{\frac{m+1}{m(m-1)}} ,
\end{equation}
where $a$ is an arbitrary positive constant. By integrating both sides of \Eq{eq:ab_diff_func_1}, we obtain
\begin{equation}\label{eq:ab_diff_func_2}
-\lambda^{\frac{m(m-1)}{m+1}} \omega F = \tfrac{m+1}{m} \omega^{\frac{3m+1}{m+1}} (F^m)_\omega ,
\end{equation}
which is equivalent to $-\frac{\lambda^{\frac{m(m-1)}{m+1}}}{m+1} \omega^{-\frac{2m}{m+1}} = F^{m-2} F_\omega$.  As $x \to -\infty$, we have $\omega(\phi,\tau) = \phi(x)/T(\tau) \to 0$. From \Eq{eq:ab_diff_func_2}, the boundary condition is shown $F_\omega(0) = 0$. Again, \Eq{eq:ab_diff_func_2} can be solved for the solution
\begin{equation}\label{eq:ab_diff_func_3}
F(\omega) = \lambda \left[ b + (\lambda\omega)^{-\frac{m-1}{m+1}} \right] ^ \frac{1}{m-1} ,
\end{equation}
where $b$ is a constant which is determined later. By substituting \Eq{eq:ab_diff_time_2} and \Eq{eq:ab_diff_func_3} into \Eq{eq:ab_diff_sol}, we obtain the transformed density
\begin{equation}\label{eq:ab_diff_func_5}
\Phi(\phi,\tau) =  \left\lbrace \frac{1}{(\tau + a)^\frac{m+1}{m}} \left[b + \frac{\phi^{-\frac{m-1}{m+1}} }{ (\tau + a)^{-\frac{1}{m}} } \right] \right\rbrace^\frac{1}{m-1} .
\end{equation}
We substitute \Eq{eq:ab_diff_func_5} into the main solution \Eq{eq:sol_1}, by doing some algebra, then we obtain 
\begin{equation}
\label{eq:main_sol_final}
u(x,t) =  
\frac{ e^{t} }{ (e^{pt} -1 + a)^\frac{1}{p} }  \left[ 1 + b\left( \frac{ e^{p x} }{ e^{pt} -1 + a } \right)^\frac{1}{p+1} \right] ^ \frac{1}{p} , 
\end{equation}
which is the implicit solution of \Eq{eq:gen_fisher_dimless}.

Our selection of the undetermined constants $a$ and $b$ are dependent on the initial- and boundary conditions. At $t = 0$, we have the initial profile as $u_{0}(x) := u(x,0) = (1/a^{1/p}) \left[ 1 + b\left( e^{p x}/a \right)^{1/(p+1)} \right]^{1/p}$. If the solution has the \textit{(semi) compact support} \cite{Biro1997, Rosenau2002, Harris2004}, the density must be vanish at an initial front position $x_{0}$, i.e. $u_{0}(x_{0}) = 0$. We then evaluate the constant $b = -( a e^{-p x_{0}} )^{1/(p+1)}$. The initial density now becomes $u_{0}(x) =  (1/a)^{1/p} \left[ 1 - e^{p(x - x_{0})/(p+1)} \right]^{1/p}$. As $x \ll x_{0}$, density must reach the constant value $\rho =  \displaystyle{\lim_{x - x_{0} \to -\infty}} u_{0}(x) = ( 1/a )^{1/p}$. That is, we have $a = \rho^{-p}$. By substituting the constant $a$ and $b$ into \Eq{eq:main_sol_final}, we obtain the exact solution
\begin{eqnarray}
\label{eq:main_sol_exact}
\lefteqn{
u(x,t) =  \frac{ \rho e^{t} }{ [\rho^{p}(e^{p t} -1) +1]^\frac{1}{p}} } \nonumber \\ && \times 
\left\lbrace  1 - \left[ \frac{ e^{p (x-x_{0})} }{ \rho^{p}(e^{p t} -1) +1 } \right]^\frac{1}{p+1}   \right\rbrace ^ \frac{1}{p} ,
\end{eqnarray}
which the initial density profile has the form
\begin{equation}
\label{eq:main_sol_exact_init} 
u_0(x) = \rho \left[ 1 - e^{\frac{p}{p+1} (x-x_{0})} \right] ^ \frac{1}{p} . 
\end{equation}
As $x \ll x_0$, the density converges to the equilibrium state $\displaystyle\lim_{t \to \infty} u(x,t) = 1$. 

We have also found another solution for \Eq{eq:gen_fisher_dimless}. In the factorized form, we rearrange \Eq{eq:gen_fisher_dimless} as $(\frac{\partial}{\partial t} - 1)u = (\frac{\partial}{\partial x} - 1)(\frac{\partial}{\partial x} + 1)u^m$. Then, it can be written in the equivalent equation 
\begin{equation}\label{eq:tranfisher_1_neg}
e^{t}\frac{\partial}{\partial t}e^{-t} u = e^{x}\frac{\partial}{\partial x}e^{-x}(e^{-x}\frac{\partial}{\partial x}e^{x} u^m) . 
\end{equation}
By omitting the details, similarly, by defining another solution $u_{-}(x,t) = e^{-\frac{x}{m}} e^t \Phi(x,t)$ for \Eq{eq:gen_fisher_dimless}, we obtained the same anomalous diffusion equation \Eq{eq:ab_diff_1} and the same temporal transform-function \Eq{eq:time_1} with the different spatial transform-function $\phi_{-}(x) =  e^{-\frac{m+1}{m}x} = \phi(-x)$. Later, it is found the relation between the two solutions actually that 
\begin{equation}\label{eq:u_}
u_-(x,t)=u(-x,t) . 
\end{equation}

It is obvious that the two solutions are mutual symmetry.  Once, it was suggested that we can construct the symmetric solution \cite{Newman1980, Biro1997, Rosenau2002} by the linear combination of
\begin{equation}\label{eq:lin_comb}
w(x,t) := u(x,t)+u_{-}(x,t) . 
\end{equation}
To see how \Eq{eq:lin_comb} applicable is, let us consider the combination of \Eq{eq:gen_fisher_dimless} of the two solutions $(u+u_-)_t = (u^{m}+u_-^{m})_{xx} + (u+u_-) - (u^{m}+u_-^{m})$. By using the binomial expansion $u^{m}+u_-^{m} = (u+u_-)^m -\sum_{k=1}^{m-1}\binom{m}{k}u^{m-k} u_-^{k}$ for any integer $m$ and $u^{m}+u_-^{m} = (u+u_-)^m  - \left[\sum_{k=1}^{\infty}\binom{m}{k} u^{m-k} u_-^{k} - u_-^m \right]$ for any fraction $m$, we obtain the approximation
\begin{equation}\label{eq:binonial_1}
u^{m}+u_-^{m} \approx (u+u_-)^m + \mathcal{O}( H ) ,
\end{equation}
where $H(x,t) = \sum_{k=1}^{m-1}\binom{m}{k}u^k(x,t) u_-^{m-k}(x,t)$ for integer $m$ and $H(x,t) = \sum_{k=1}^{\infty}\binom{m}{k} u^{m-k}(x,t) u_-^{k(x,t)} - u_-^m(x,t) $ for fraction $m$. Using  \Eq{eq:binonial_1}, then we obtain 
\begin{equation}\label{eq:gen_fisher_comb_2}
w_t = (w^{m})_{xx} + w - w^{m} + \mathcal{O}(H_{xx}-H) .
\end{equation}
Therefore, the linear combination $w(x,t)$ in \Eq{eq:lin_comb} can be the solution of \Eq{eq:gen_fisher_dimless} with the correction $\mathcal{O}(H_{xx}-H)$. For any $p>0$, using the binomial expansion \Eq{eq:binonial_1} and modifying the initial condition, we have the symmetric solution
\begin{eqnarray}
\label{eq:main_sol_exact_sym}
\lefteqn{
w(x,t) \approx  \frac{ \rho e^{t} }{ [\rho^{p}(e^{p t} -1) +1]^\frac{1}{p}} } \nonumber \\ && \times
\left\lbrace  1 - \frac{\Cosh{\frac{p}{p+1} (x-x_{0})} }{ \left[\rho^{p}(e^{p t} -1) +1 \right]^\frac{1}{p+1}  }  \right\rbrace ^ {\frac{1}{p}} ,
\end{eqnarray}
where $\Cosh(x+k) := \frac{1}{2}(e^{x+k}+e^{-x+k}) = e^{k}\cosh x$ is the modified hyperbolic function. It is found that the symmetric initial density $w_0(x) := w(x,0)$ is given by
\begin{equation}
\label{eq:main_sol_exact_sym_init}
w_0(x) \approx  \rho \left[  1 - \Cosh{\frac{p}{p+1} (x-x_{0})}   \right] ^ {\frac{1}{p}} .
\end{equation}
Here, we found the special forms of the symmetric initial density $w_0(x)$. If $x_0$ is sufficiently small $x_0 \ll 1$, the initial density forms the delta-like profile $w_0(x)\approx \rho\delta(x)$.  If $x_0$ and $p$ are sufficiently large, $x_0 \gg 1$ and $p \gg 1$, the  initial density forms the rectangular profile $w_0(x)\approx \rho\sqcap(x)$. 

\begin{figure}
\includegraphics[width=\columnwidth]{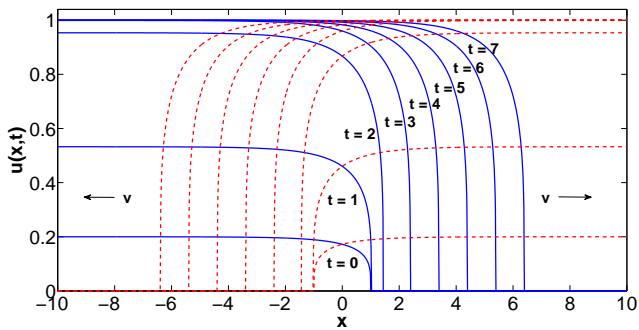}
\caption{\label{fig:evolution}
The spatiotemporal evolution of the density profile in the case $p=4$ with the initial conditions $\rho=0.2$ and $x_{0}=1$. The solid lines represent $u(x,t)$ and the dashed lines represent $u_-(x,t)=u(-x,t)$.
}
\end{figure}

\begin{figure}
\includegraphics[width=\columnwidth]{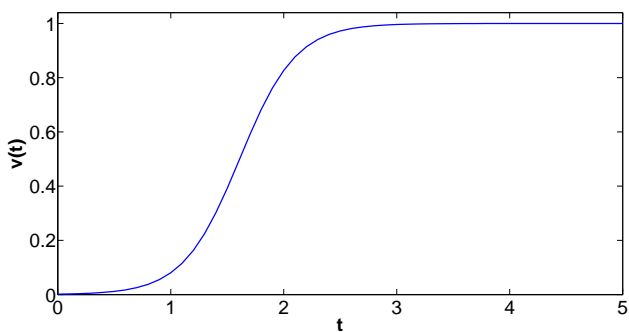}
\caption{\label{fig:velocity}
The spreading velocity $v(t)$ of $u(x,t)$ corresponding to \Fig{fig:evolution}. The initial speed is $v(0)=\rho^{4}=0.0016$ and the transition time is $t^{\prime} \approx -\ln \rho = 1.61$.
}
\end{figure}

\section{\label{sec:discussions}Discussions}
The evolution of solution $u(x,t)$ in \Eq{eq:main_sol_exact} and its corresponding $u_-(x,t)=u(-x,t)$ are illustrated in \Fig{fig:evolution}. It has seen that $u(x,t)$ is spreading to the right whereas $u_{-}(x,t)$ is spreading to the left, with velocity $\pm v(t)$ respectively. The solutions grow to the saturated density profile with the maximum value $1$ by initiating $\rho < 1$. In other hand, the solutions decay to the saturated density profile with the maximum value $1$ by initiating  $\rho > 1$ (has not been illustrated in the figure here).

\subsection{Spreading velocity and transition time}
As the solution \Eq{eq:main_sol_exact} has the compact support, the density falls to zero $u(r,t) = 0$  after the front position $r(t)$. Consequently, we obtain the front position for the right-spreading solution only: 
\begin{equation}\label{eq:front_postion}
r(t) = x_{0} + \frac{\ln ( \rho^p(e^{pt} -1) + 1 ) }{p} .
\end{equation} 
Then, the spreading velocity is calculated as
\begin{equation}\label{eq:front_speed_1}
v(t) = \frac{\Dif r(t)}{\Dif t} = \frac{ \rho^p e^{pt} }{ \rho^p(e^{pt} -1) + 1 } .
\end{equation}
The spreading velocity versus time is illustrated in \Fig{fig:velocity}. If the initial density constant is not equal to the equilibrium point $\rho \neq 1$, it is seen that the dynamics of the systems can be separated at least in two regimes. At the early stage, the systems spread rapidly from an initial speed $v(0) = \rho^{p}$. After the transition time $t^{\prime}$, the systems spread with the constant the spreading speed $c = \displaystyle{\lim_{t \to \infty}} v(t) = 1$ at the late stage. Therefore, the transition time $t^{\prime}$ can be estimated from \Eq{eq:front_speed_1} by using the fact $\rho^p(e^{p t^{\prime}} -1) \approx 1$. Then, we obtain
\begin{equation}\label{eq:transition_t}
t^{\prime} \approx  \ln (1+\frac{1}{\rho^p})^\frac{1}{p} ,
\end{equation}
which is exactly depended on the initial density constant $\rho$. If the initial density constant is much greater than the equilibrium density $\rho \gg 1$, the transition time is expected $t^{\prime} \approx 0$. Otherwise, if the initial density constant is much lesser than the equilibrium density $\rho \ll 1$, transition time is expected $t^{\prime} \approx  -\ln \rho$. 

\subsection{Traveling wave solutions}
At the large time scale $t \gg t^{\prime}$, the solution \Eq{eq:main_sol_exact} becomes 
\begin{equation}\label{eq:travel_wave_1}
\widetilde{u}(x-ct) = \left[ 1 - \frac{e^{\frac{p}{p+1}(x - t - x_{0})}}{\rho^{\frac{p}{p+1}}} \right] ^ \frac{1}{p} ,
\end{equation}
which is known as the traveling wave solution where $c = 1$ is the front speed. When converting to the physical dimension, the front speed obtained here $c=\sqrt{\kappa\alpha/(p+1)}$ is comparable to the minimum value that the traveling wave emerged \cite{Newman1980, Newman1983, Murray2002, Petrovskii2006}. If the initial density constant is equal to the equilibrium density $\rho = 1$, the traveling wave emerges from $t=0$ without the transition time. Similarly, at the large time scale $t \gg t^{\prime}$, \Eq{eq:main_sol_exact_sym} develops to the symmetric traveling wave solution
\begin{equation}
\label{eq:main_sol_exact_sym_wave}
\widetilde{w}(x-c t) \approx  
\left[ 1 - \frac{\Cosh{\frac{p}{p+1} (x-t-x_{0})} }{\rho^\frac{p}{p+1}}  \right]^{\frac{1}{p}} ,
\end{equation}
which is the compact expanding wave with the expanding speed $c=1$. We now have the traveling wave solutions either spreading to the right or to the left and the symmetric traveling wave solution spreading both direction, as shown by \cite{Kamin2004}. 

We point out that, at large time scale $t \gg t^{\prime}$, the solution \Eq{eq:ab_diff_func_5} forms the self-similar structure $\Phi(\phi,\tau) \approx \frac{1}{\tau^\beta} F(\frac{\phi}{\tau^\beta})$, where $\beta = \frac{m+1}{m(m-1)}$, which can be classified as the intermediate asymptotics of the second kind as described in \cite{Barenblatt1972, Barenblatt1996}. The traveling wave solution \Eq{eq:travel_wave_1} $\widetilde{u}(x - t - x_{0})$ is connected to the self-similar solution by the transformations $\widetilde{u} \to e^{x/m} e^{t} \Phi$, $\tau \to e^{(m-1)t}$, and $\phi \to e^{(m+1)(x-x_0)/m}$.

\subsection{Biological population dynamics}
In the term of biological system, the solution \Eq{eq:main_sol_exact_sym} captures the qualitative features of the spatiotemporal pattern formation bacterial colonies \cite{Kawasaki1997, Golding1998, Ben-Jacob2000}. The bacteria are inoculated from an initial density profile then the bacteria reproduce until reach the saturated number due to the limitation of nutrient while the bacteria expand the colony to localize the higher nutrient region. Surprisingly, this process has reflected the self-similar structure.

In the opposite of the logistic law, $\alpha < 0$, \Eq{eq:gen_fisher_dimless} becomes $u_t = (u^{m})_{xx} + u^{m} - u$, in the dimensionless form. This equation is exactly the porous medium equation with source and sink term. We have found that by the transformations $t \to -t$ and $x \to ix$, absolutely, the symmetric solution \Eq{eq:main_sol_exact_sym} would be applicable. Since the traveling wave pattern has not emerged in this system, the further analysis on this equation is not scope of this paper.

\section{\label{sec:conclusions}Conclusions}
We studied the spatiotemporal pattern formation in a density-dependent reaction-diffusion equation which is known as the extension of the Fisher or Kolmogoroff-Petrovsky-Piscounoff equation. The exact self-similar solutions with the compact support were found. The solutions converge to the well-known traveling wave solutions at $t\to\infty$. Therefore, the connection between the self-similar solutions and the traveling wave solutions has been shown clearly from the analytical solutions. 

\begin{acknowledgments}
K. Khompurngson would like to acknowledge the Centre of Excellence in Mathematics (Thailand) for the partial financial support. 
\end{acknowledgments}


\bibliography{DDRDE_ref}

\end{document}